\documentclass{PoS}

\title{Improved measurement of parton distribution functions and $\alpha_s(M_Z)$ with the LHeC}

\ShortTitle{High Precision DIS with the LHeC}

\author{\speaker{Amanda Cooper-Sarkar}\thanks{On behalf of LHeC Collaboration}\\
        Oxford University\\
        E-mail: \email{amanda.sarkar@cern.ch}}

\abstract{The potential of the LHeC, a future electron-proton collider, 
for precision Deep Inelastic Scattering measurements is reviewed with 
particular emphasis on the reduction of uncertainties on the 
parton distribution functions (PDFs) of the proton and on the measurement of $\alpha_s(M_Z)$. The interpretation of 
possible Beyond Standard Model (BSM) signals at the LHC is crucially dependent 
on precise knowledge of the predictions of the Standard Model (SM) and 
the uncertainties on PDFs and $\alpha_s(M_Z) $are a limiting factor. 
The LHeC project, running in 
parallel with later stages of LHC running, would provide much improved 
precision on the PDFs as compared to the precision expected from LHC data 
alone.}

\FullConference{XXIV International Workshop on Deep-Inelastic Scattering and Related Subjects\\
		11-15 April, 2016\\
		DESY Hamburg, Germany}

\begin{document}

The LHeC project proposes a Large Hadron-Electron Collider using  
a Linac-Ring configuration with electrons of 50-100 GeV colliding 
with 7 TeV protons in the LHC tunnel, designed such that e-p collisions can 
operate synchronously with p-p. The details of the accelerator and the detector
are covered in other contributions to this conference. 
This talk focusses on Deep Inelastic Scattering and low-x physics. Higgs, 
BSM physics and e-A collsions are covered in other talks. Further details 
may be found in the Conceptual Design Report (CDR)~\cite{cdr}.

The LHeC reperesents an increase in the kinematic reach of Deep Inelastic 
Scattering and an increase in the luminosity. 
This allows a potential increase in the precision of parton
 distributions in the kinematic region of interest for the detailed understanding 
of BSM physics at the LHC. It also allows the exploration of a kinematic region at 
low-x  where we learn more about QCD- is there a need for resummations beyond 
DGLAP, or even for non-linear evolution and gluon saturation? 

If a 100TeV Future Circular Collider (FCC) is built for hadron-hadron 
collisions, FCChh, this will extend the kinematic region for BSM physics 
further and it is worth contemplating an electron-hadron collider, FCCeh, 
which would provide a much more precise knowldege of parton distributions 
in this new kinematic region as well as exploration of QCD at even lower $x$.
 
Deep Inelastic Scattering is the best process to probe proton structure. The 
Neutral 
Current Cross Sections meaure the sea quarks and access the gluon via the 
scaling 
violations and the longitudinal structure function. The Charged Current processes 
give information on flavour separated valence quarks and the difference between 
the Neutral Curremt $e^+$ and $e^-$ distributions probes the valence quarks 
via the $\gamma-Z$ interference term.

To study the potential of the LHeC a scenario with 50 GeV electrons on 7 TeV protons 
with 50 fb$^{-1}$ luminosity is simulated. The kinematic region accessed is 
$0.000002 < x< 0.8$ and $2 < Q^2 <100,000$GeV$^2$. Uncorrelated and correlated 
systematic errors 
are simulated using our knowledge of dominant sources such as the electron 
and hadron energy scales, angular resolution and photoproduction background, 
based on experience with the 
H1 detector, see the LHeC CDR~\cite{cdr} for details.

\begin{figure}[tbp]
\begin{center}
\begin{tabular}{cc}
\includegraphics[width=0.4\textwidth]{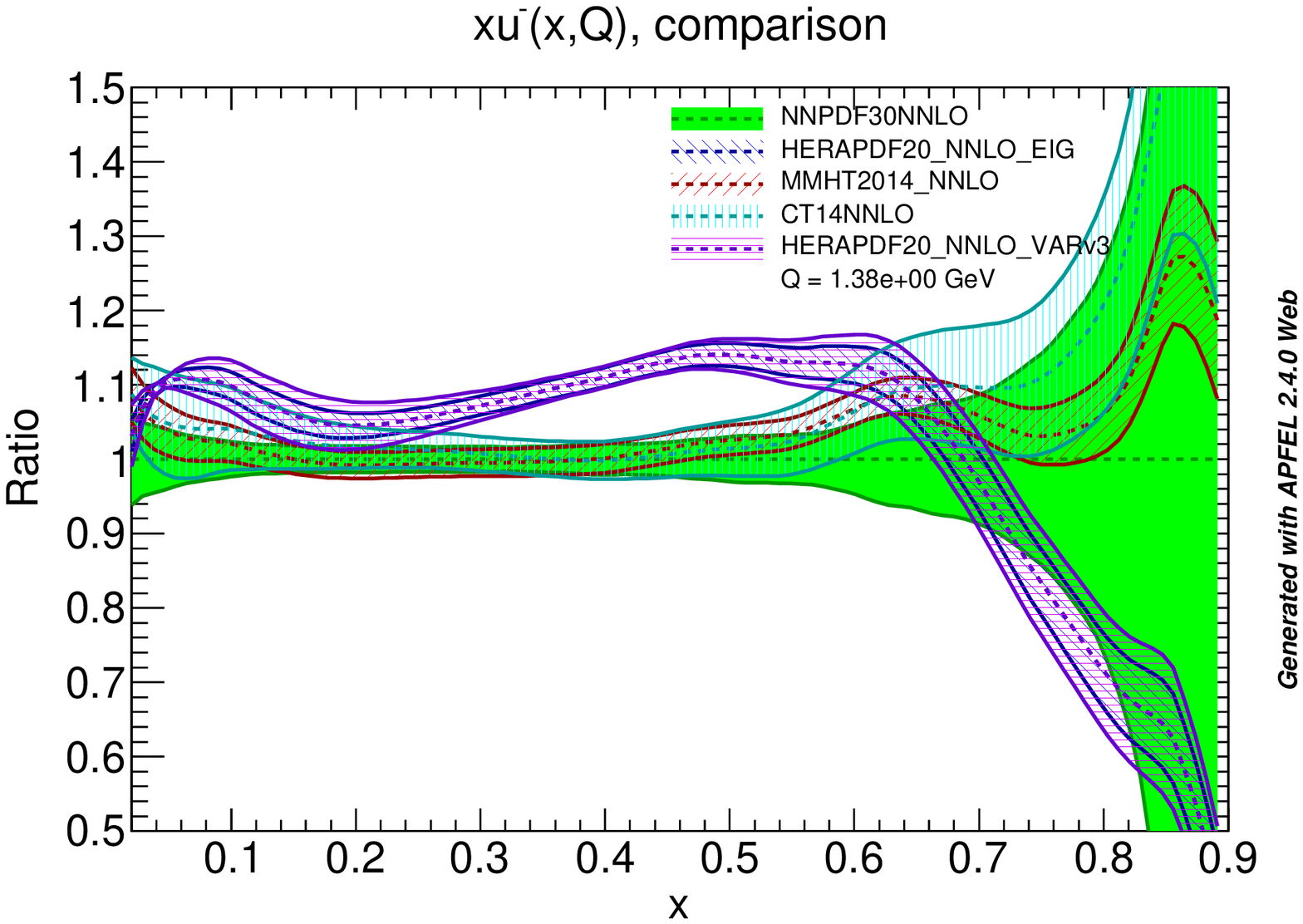} &
\includegraphics[width=0.4\textwidth]{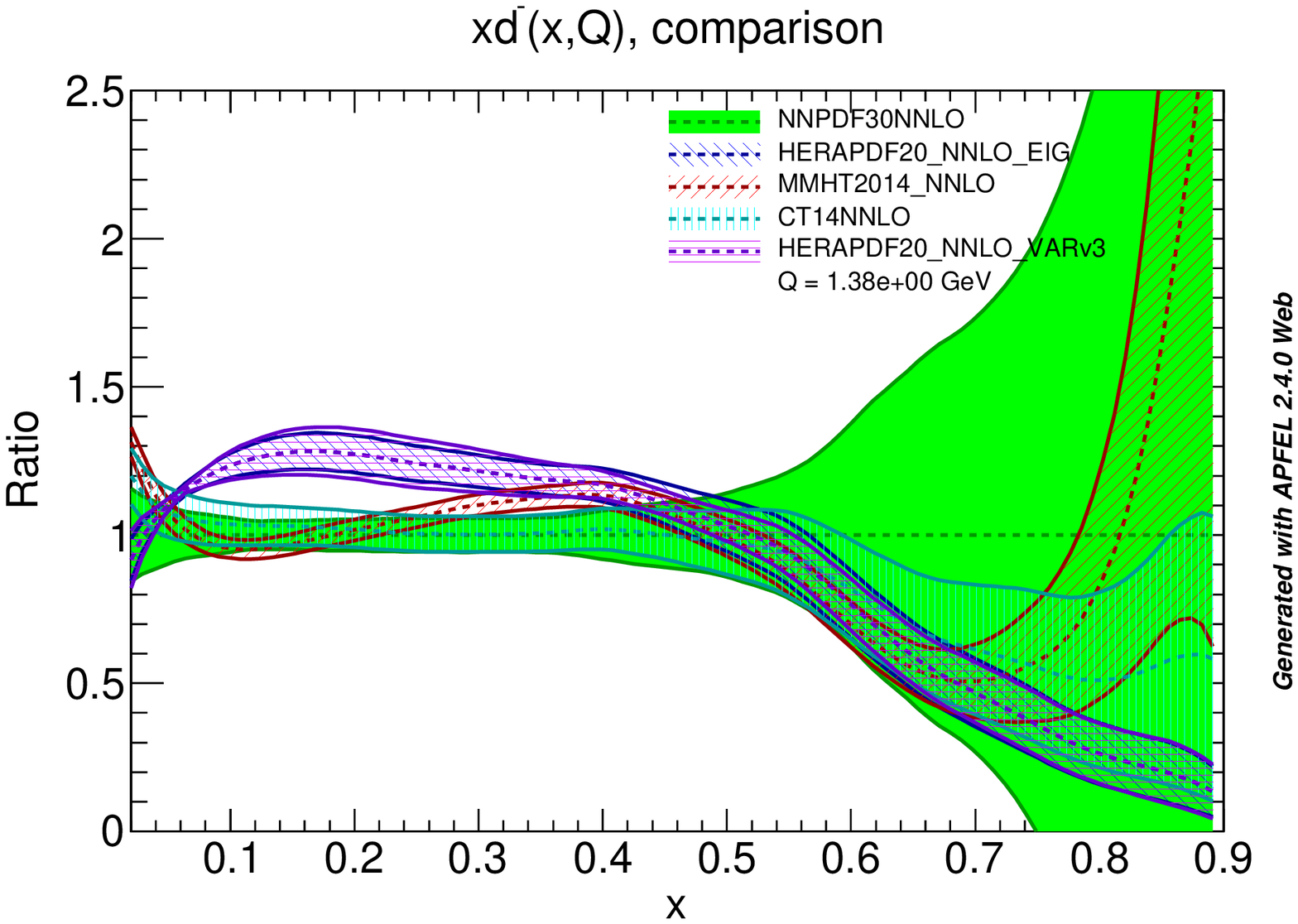}
\end{tabular}
\end{center}
\caption {Up and down valence distributions for various modern 
PDF sets in ratio to NNPDF3.0.   
}
\label{fig:valence1}
\end{figure} 
\begin{figure}[tbp]
\begin{center}
\includegraphics[height=0.3\textheight]{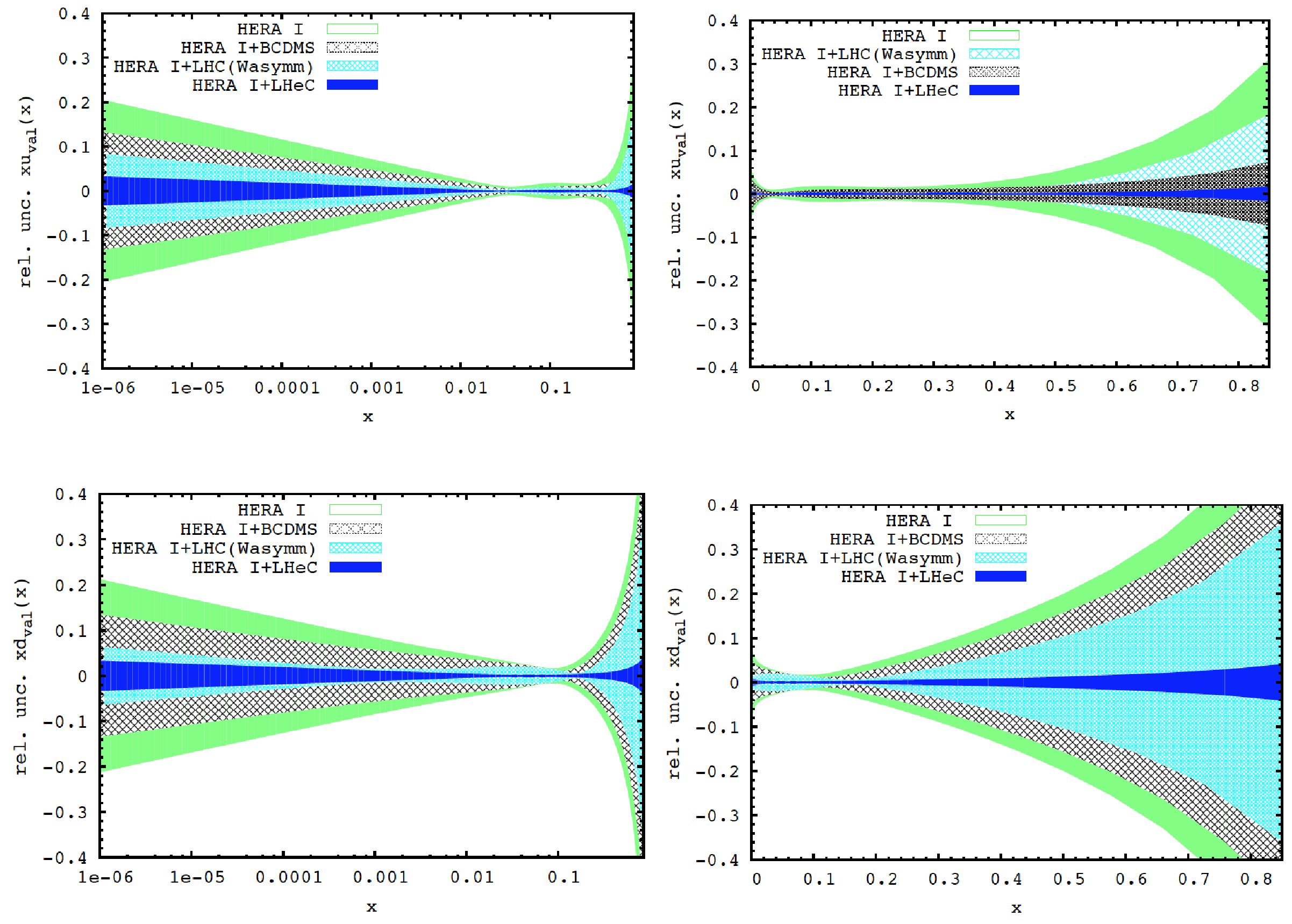}
\end{center}
\caption { The PDF 
uncertainty on the valence distributions from a fit to just HERA-I data, 
HERA-I+BCDMS data, HERA-I+LHC W-asymmetry data and HERA-I+LHeC pseudo-data. 
Log and linear axes illustrate the improved level 
of uncertainty at low and high-$x$, respectively.
}
\label{fig:valence2}
\end{figure} 
In Fig.~\ref{fig:valence1} the current level of our knowledge of valence distributions is shown, 
comparing various modern PDF sets in ratio to NNPDF3.0NNLO. In Fig.~\ref{fig:valence2} the potential improvement in uncertainty from LHeC data 
is shown by comparing the uncertainties on the valence PDFs from a fit 
to just HERA-I combined data~\cite{h1zeus:2009wt} to a fit to these data 
plus LHeC pseudo-data. Fits to HERA plus BCDMS
fixed target data and HERA plus LHC W-asymmetry data are also illustrated but 
these do not 
bring such a dramatic reduction in uncertainty, even when current LHC data 
have their uncertainties reduced to reflect our best estimate of the 
ultimate achievable accuracy. 

\begin{figure}[tbp] 
\begin{center}
\begin{tabular}{cc}
\includegraphics[width=0.4\textwidth]{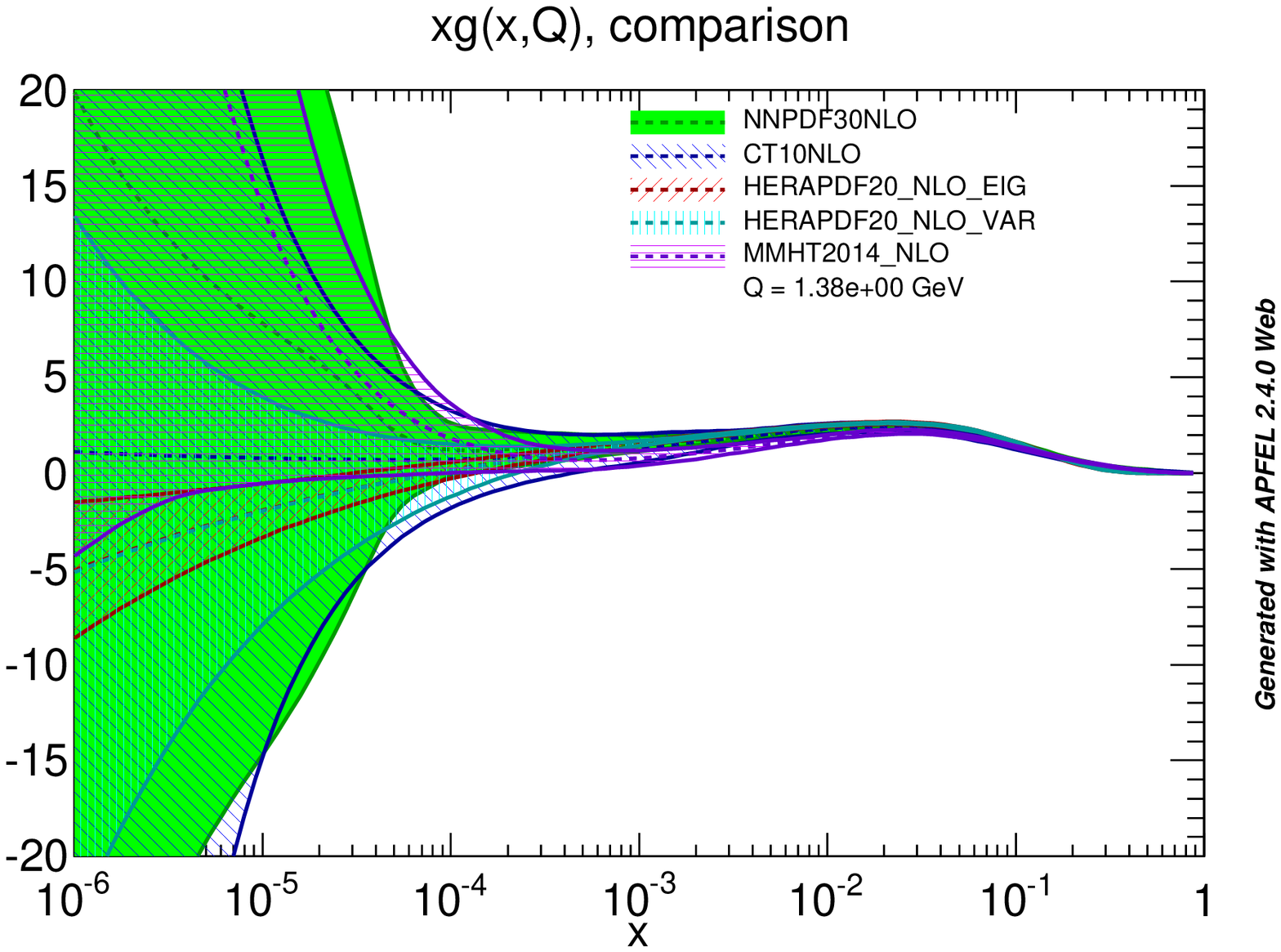} &
\includegraphics[width=0.4\textwidth]{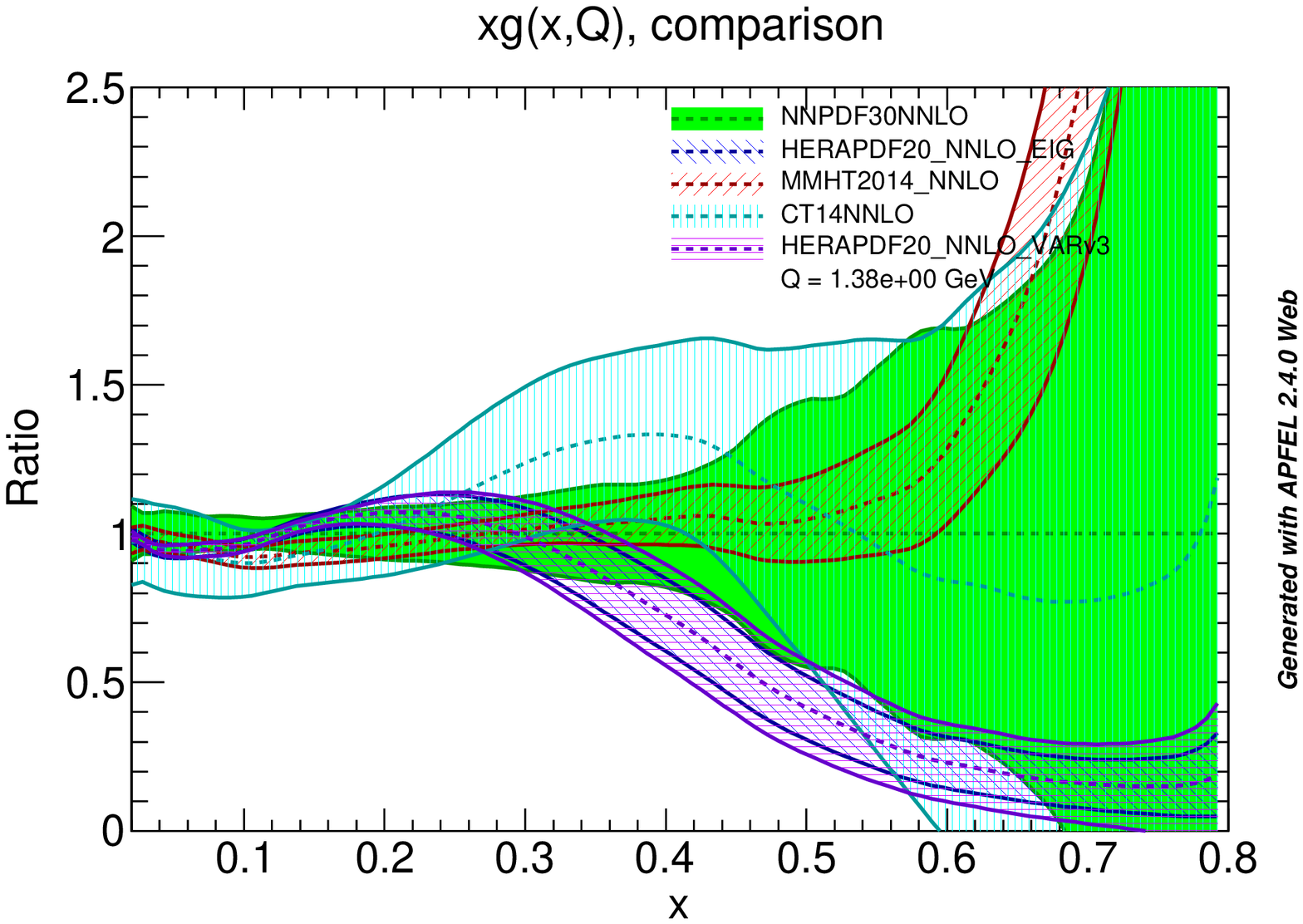}
\end{tabular}
\caption{Gluon distributions for various modern 
PDF sets in ratio to NNPDF3.0. 
Log and linear axes illustrate the level 
of uncertainty at low and high-$x$, respectively.  }
\label{fig:seaglu1}
\end{center}
\end{figure}
\begin{figure}[tbp] 
\begin{center}
\includegraphics[width=0.8\textwidth]{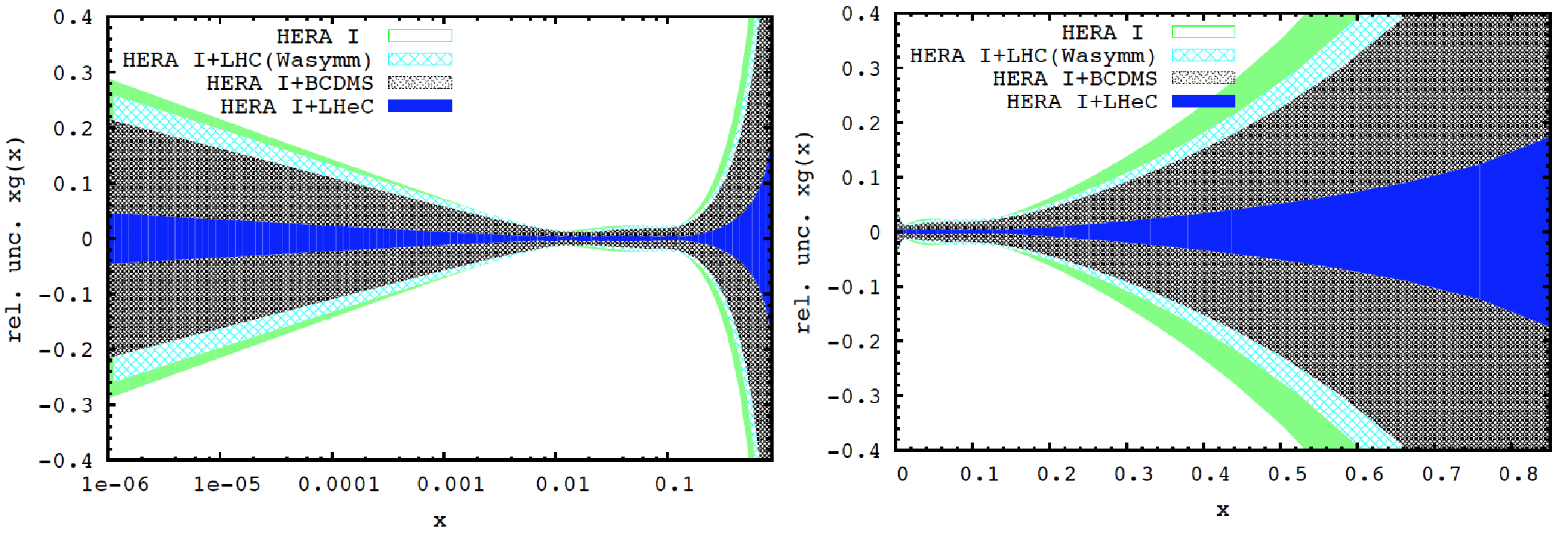} 
\caption{ The PDF 
uncertainty on the gluon distribution from a fit to just HERA-I data, 
HERA-I+BCDMS data, HERA-I+LHC W-asymmetry data and HERA-I+LHeC pseudo-data. }
\label{fig:seaglu2}
\end{center}
\end{figure}
In Fig.~\ref{fig:seaglu1} the current level of our knowledge of the gluon 
distribution is shown, 
comparing various modern PDF sets in ratio to NNPDF3.0. Log and linear axes illustrate the current level 
of uncertainty at low and high-$x$, respectively. Fig.~\ref{fig:seaglu2} shows the expected level of improvement from LHeC data.
 
\begin{figure}[tbp]
\begin{tabular}{cc}
\includegraphics[width=0.5\textwidth, angle=0]{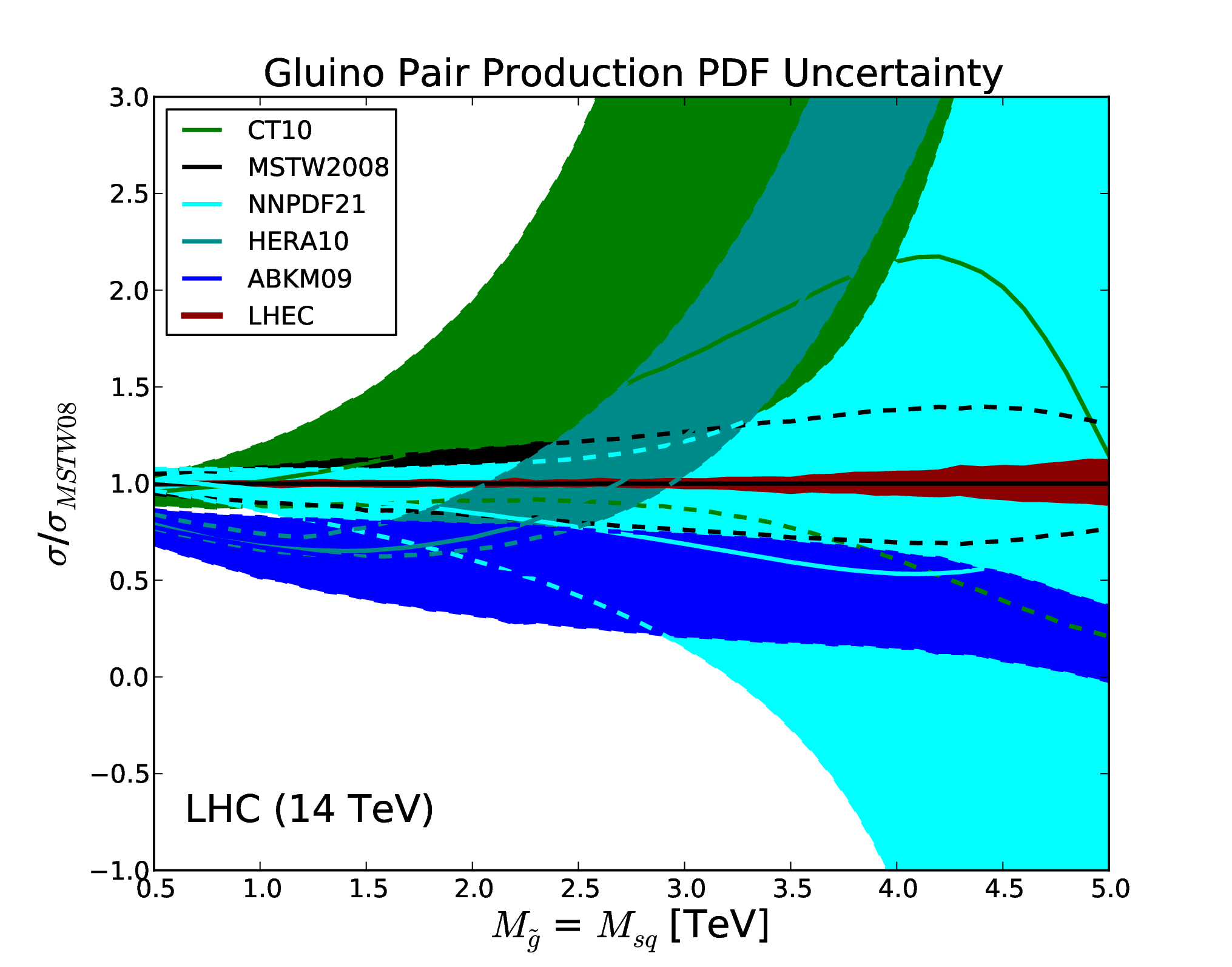}
\includegraphics[width=0.37\textwidth, angle=0]{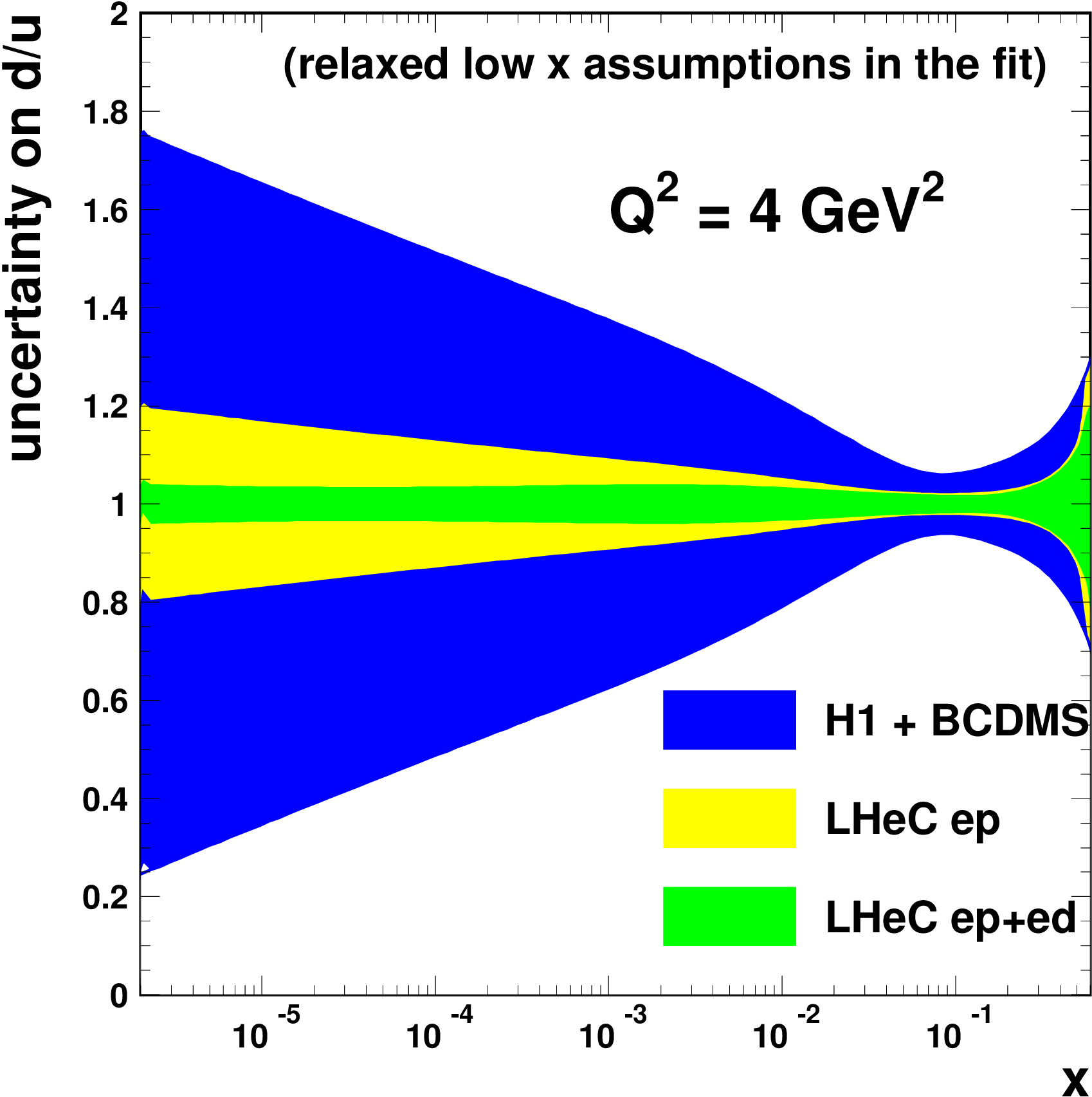}
\end{tabular} 
\caption {Left-hand side: Gluino pair production cross-section for various 
PDFs in ratio to MSTW2008, as calculated in NLO SUSY QCD assuming squark mass 
degeneracy and equality of squark and gluino masses. Right-hand side: 
PDF uncertainty on the $d/u$ ratio, relaxing the assumption $\bar{d}=\bar{u}$ 
at low $x$, for current data and after LHeC pseudo-data is used from 
both $ep$ and $eD$ runs.  
}
\label{fig:gluino}
\end{figure}

As an example of the importance of such precision at high $x$, 
Fig.~\ref{fig:gluino} left-hand side, shows a plot of the PDF uncertainty on the gluino pair production cross 
section as a function of energy, from current PDFs and from the projected 
post-LHeC PDF. Such gain in PDF precision will be necessary to exploit the 
gain in experimental precision of future searches for gluinos when the LHC
luminosity is increased from $0.3$ab$^{-1}$ to $3$ab$^{-1}$.

\begin{figure}[tbp]
\begin{center}
\includegraphics[width=0.5\textwidth, angle=-90]{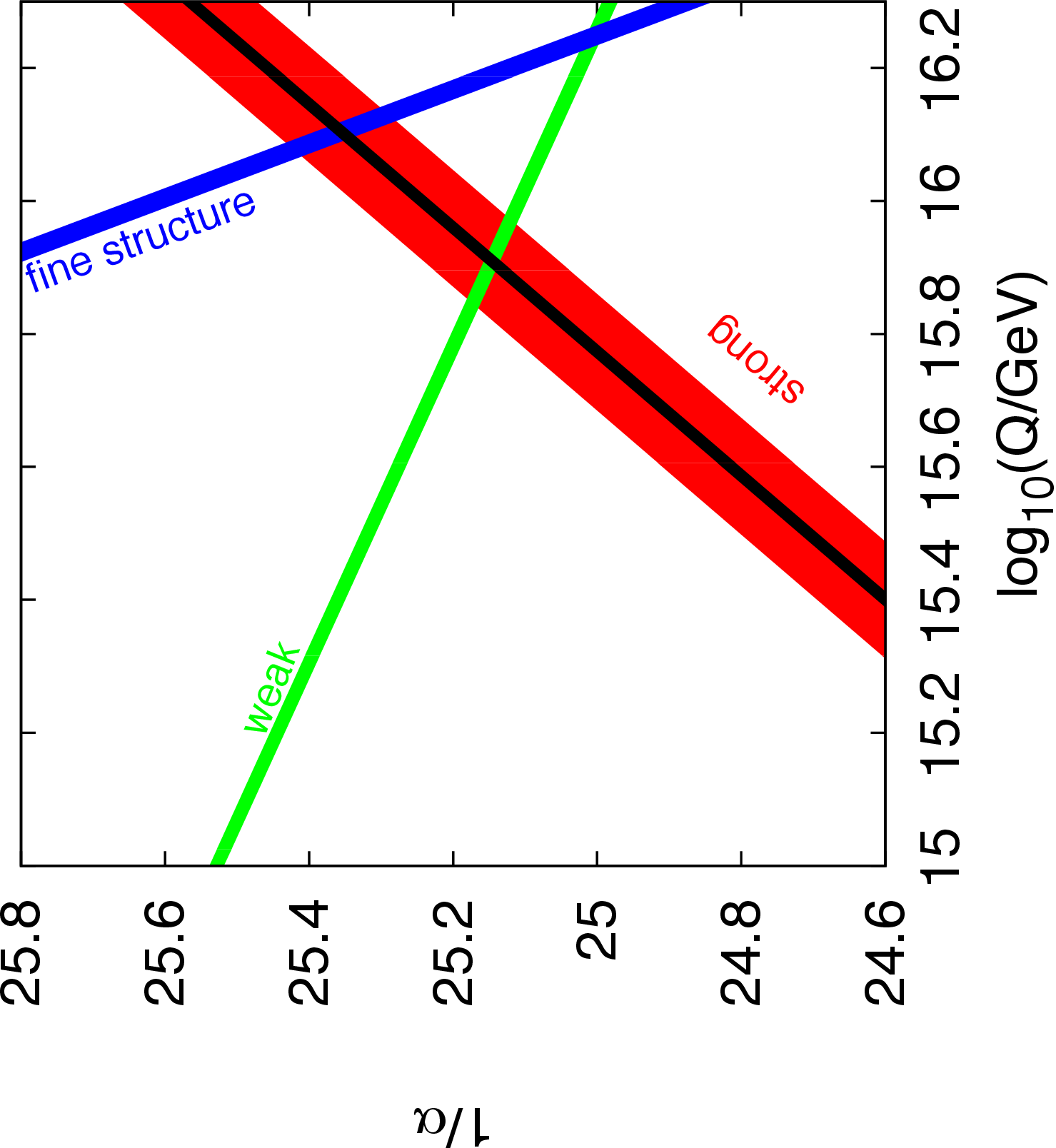}
\end{center}
\caption { 
Extrapolation of the coupling constants ($1/\alpha$) to the GUT scale in 
MSSM as predicted by SOFTSUSY. The width of the red line shows the uncertainty 
on the current world average of $\alpha_s(M_Z)$ and width of the black-line 
shows the projected accuracy of an LHeC measurement.  
}
\label{fig:alphas}
\end{figure}

The uncertainty on $\alpha_s(M_Z)$ is also important for many 
BSM cross sections. 
The LHeC can deliver per-mille accuracy on $\alpha_s(M_Z)$ and this will be a 
strong constraint on Grand Unified Theories which predict where the couplings 
unify~\cite{laycock}. This is illustrated in Fig.~\ref{fig:alphas}.

\begin{figure}[tbp]
\begin{center}
\includegraphics[width=0.6\textwidth, angle=90]{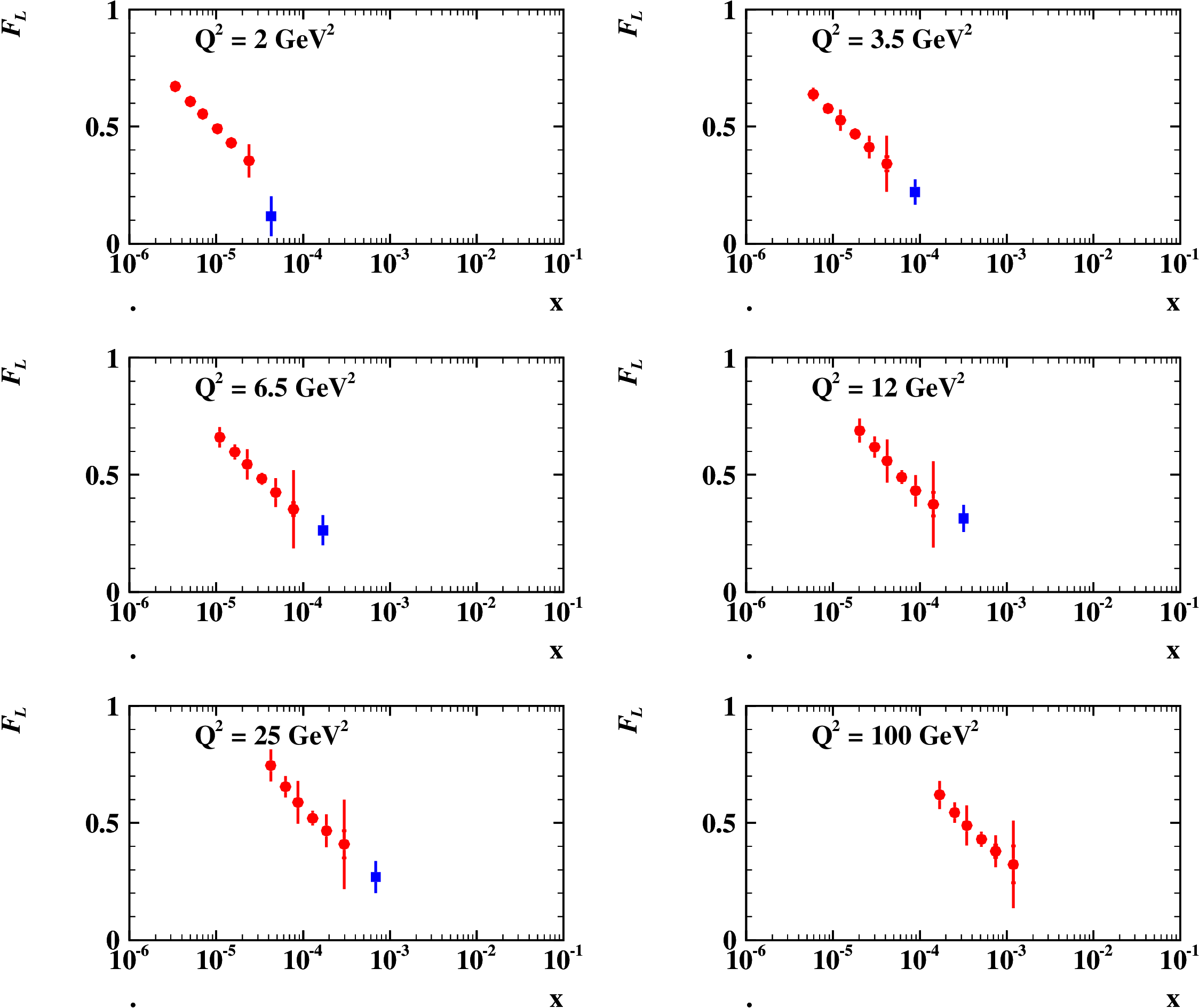} 
\end{center}
\caption {The current measurements of $F_L$ from HERA data (in blue) compared to projected measurements of $F_L$ at the LHeC (in red).
}
\label{fig:fl}
\end{figure}

Turning now to the low-$x$ region. 
Fig.~\ref{fig:seaglu1} shows that HERA sensitivity stops at $x >5\times 10^{-4}$ 
whereas the LHeC can probe down to 
$x\sim 10^{-6}$, see Fig.~\ref{fig:seaglu2}. 
Thus one can better explore the low-x region where DGLAP evolution may 
need to be supplemented by $ln(1/x)$resummation (BFKL resummation) and one 
may enter 
into a kinematic regime where non-linear evolution is required, possibly leading to 
gluon saturation.
In DGLAP based QCD fits we get the gluon from the scaling violations at low-$x$,
$dF_2/dln(Q^2)\sim P_{qg} xg(x,Q^2)$. 
The shape of the gluon extracted may be 
incorrect if the splitting function $P_{qg}$ needs modification. 
To check this one 
can measure other gluon related quantities like the longitudinal 
structure function 
$F_L$, which is gluon dominated at low $x$, $F_L(x,Q^2)\sim xg(2.5x,Q^2)$ at LO. 
Unfortunately the final $F_L$ measurements from HERA do not have sufficient 
kinematuic range or sufficient accuracy to challenge DGLAP. 
However, Fig.~\ref{fig:fl} compares 
current measurements with the projected LHeC measurements of $F_L$, 
which should be discriminating.

Further low-x studies include relaxing the conventional assumption, 
used in many PDFs, 
that $\bar{u}=\bar{d}$ at low-$x$. The right-hand side of 
Fig.~\ref{fig:gluino} shows PDF uncertainties 
on the $d/u$ ratio with this constraint relaxed, and compares current levels 
of uncertainty with the projections from LHeC pseudo-data. Further improvement 
could be achieved with LHeC eD data.

\begin{figure}[tbp]
\begin{tabular}{cc}
\includegraphics[width=0.45\textwidth, angle=0]{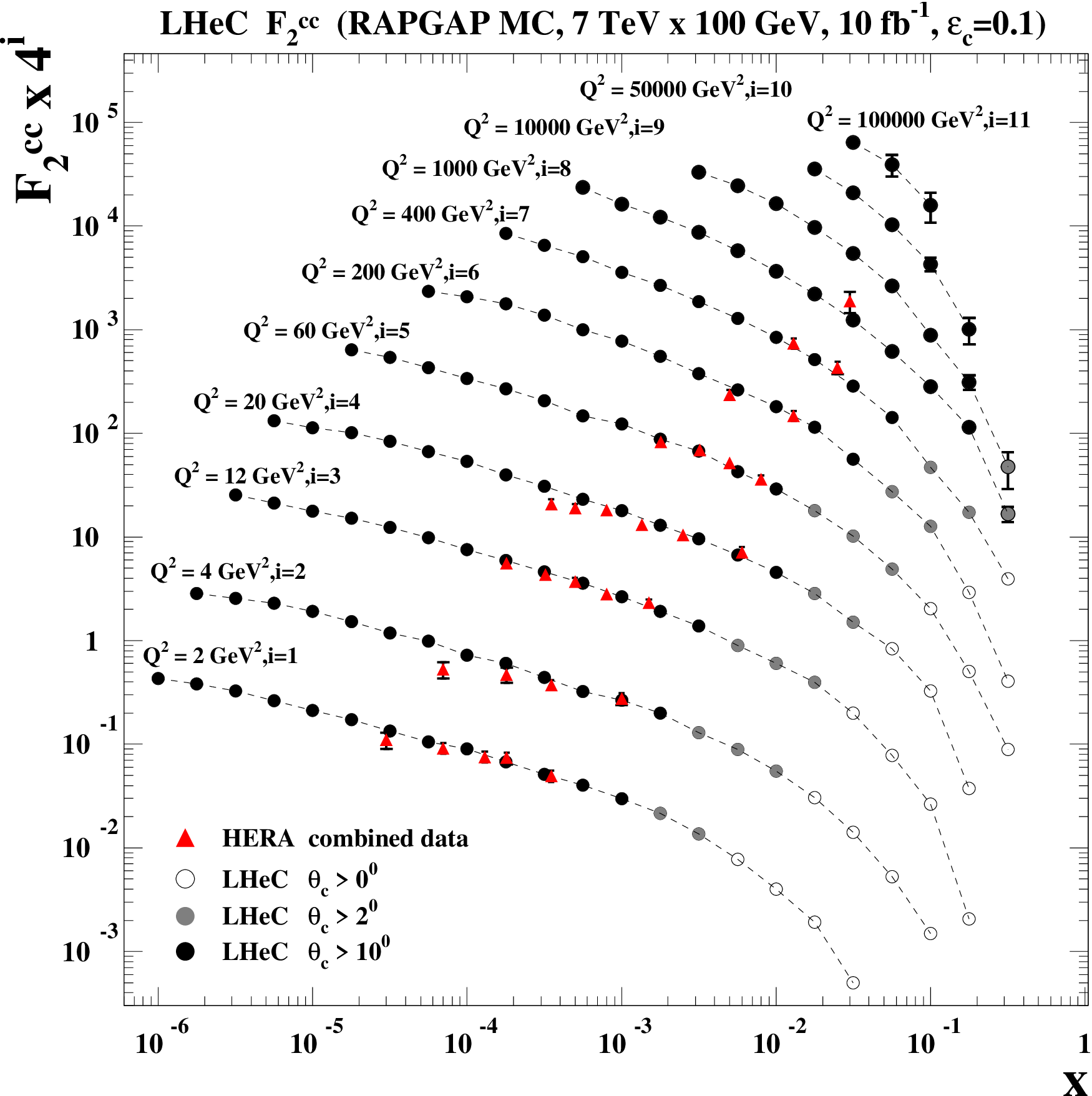} &
\includegraphics[width=0.45\textwidth, angle=0]{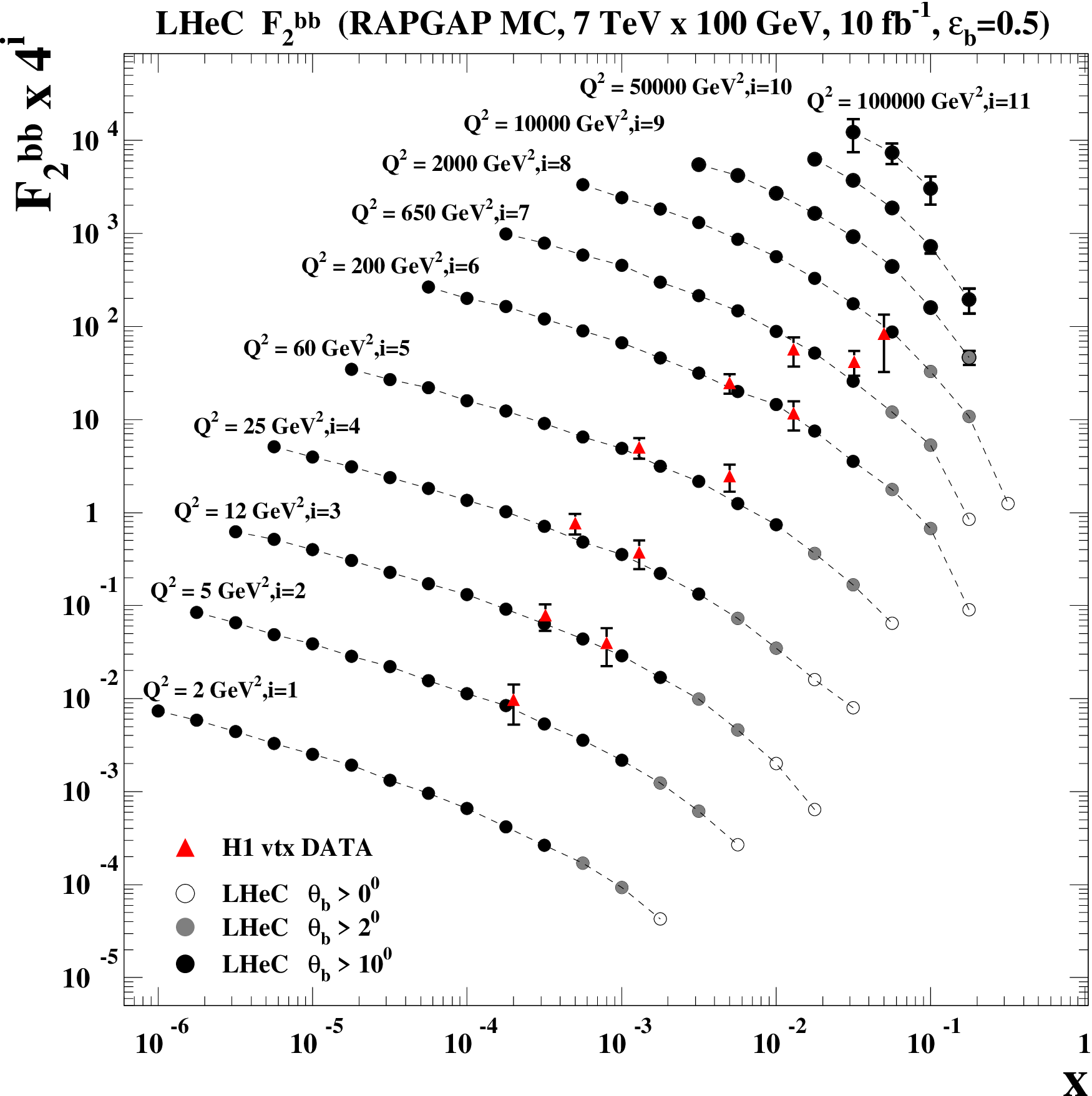}
\end{tabular}
\caption {Current measurements of $F_2^{c\bar{c}}$ (left) and $F_2^{b\bar{b}}$ 
(right) from HERA (in red) compared to projected measurements at the LHeC (in black).
}
\label{fig:hq}
\end{figure}

LHeC data will also allow us to increase our knowledge of the heavy flavour 
partons because of the higher cross sections, higher luminosity and 
new generation of Silicon detectors.
Fig.~\ref{fig:hq} shows projected measurements of $F_2^{c\bar{c}}$ and $F_2^{b\bar{b}}$ compared to present measurements. Measurements of the strange PDF 
could also be performed through charm tagging in CC events. If an FCCeh option is explored then it should also be possible to measure top PDFs produced 
through the $W b \to t$ processe.

In summary the LHeC would allow improvement in the precision of PDF determinations both 
at low $x$ and at high $x$. Improvement at high $x$, together with improved 
precision in the determination of $\alpha_s(M_Z)$ which is also expected at 
the LHeC, would allow us to predict 
BSM cross sections with sufficent accuracy to distinguish between different 
explanations of new physics phenomena. At low $x$ it has long been expected 
that extension of the conventional QCD DGLAP resummation is necessary to 
explain the data, but distinguishing between different possible scenarios such 
as BFKL resummation, non-linear evolution or the onset of gluon saturation, 
has not been possible. The improvement in accuracy at low $x$ at the LHeC 
would allow such discrimination.

\end{document}